# Kinetic and Cyber

Alexander Kott, Norbou Buchler, Kristin E. Schaefer
US Army Research Laboratory

## 1. Introduction

Although a fairly new topic in the context of cyber security, situation awareness (SA) has a far longer history of study and applications in such areas as control of complex enterprises and in conventional warfare. Far more is known about the SA in conventional military conflicts, or adversarial engagements, than in cyber ones. By exploring what is known about SA in conventional – also commonly referred to as kinetic – battles, we may gain insights and research directions relevant to cyber conflicts. For this reason, having outlined the foundations and challenges on CSA in the previous chapter, we proceed to discuss the nature of SA in conventional (often called kinetic) conflict, review what is known about this kinetic SA (KSA), and then offer a comparison with what is currently understood regarding the cyber SA (CSA). We find that challenges and opportunities of KSA and CSA are similar or at least parallel in several important ways. With respect to similarities, in both kinetic and cyber worlds, SA strongly impacts the outcome of the mission. Also similarly, cognitive biases are found in both KSA and CSA. As an example of differences, KSA often relies on commonly accepted, widely used organizing representation – map of the physical terrain of the battlefield. No such common representation has emerged in CSA, yet.

### 1.1 The Transition from a Conventional to a Virtual Battlefield

The dynamics of conflict continue to evolve over time and are historically punctuated by rapid technological advancements. The grinding attrition of industrial-age conflict of the past century, whereby interaction occurred face-to-face, is currently giving way to information-age conflict (Moffat,

2006). For reference, some key characteristics of the prior industrial-age and the current information-age are presented in Figure 1. Current Information Age conflicts encompass conventional and virtual battlefields, with perhaps an increasing emphasis on the latter.

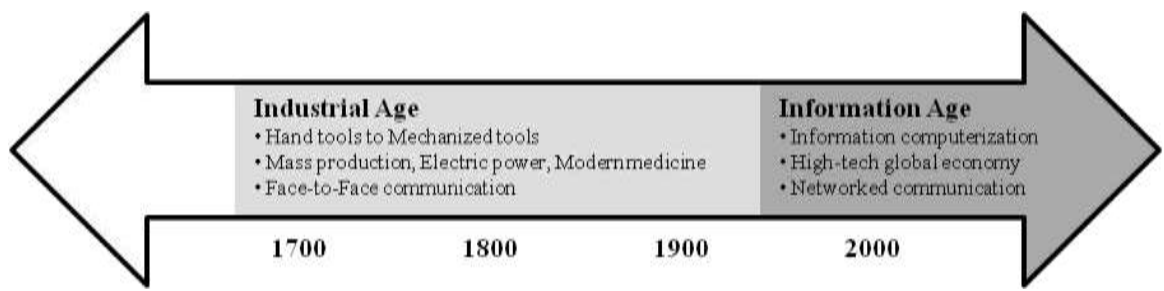

Figure 1. Characteristics of the Industrial Age to the Information Age.

In our view, the Information Age battlefield is defined by the rise of networked forms of organization. In a networked organization, the number of potential collaborators is virtually limitless, as is the availability of information. Operating in such a broadly collaborative and information-rich environment confers unprecedented advantages to a military organization (National Research Council, 2005). For instance, the transformation of U.S. and NATO countries in the late 1990s and early 2000s to networked forms of organization has given rise to large, interacting, and layered networks of Mission Command personnel communicating and sharing information within and across various command echelons as well as across joint, interagency, intergovernmental, and multinational seams and boundaries. Our dependency upon networked organizations has the consequence that warfare is no longer limited to the physics of the conventional battlefield. Increasingly, conflicts are waged purely across networks in virtual cyberspace.

A departure for our comparison is to understand the domain characteristics of kinetic and cyber operations highlighted in Table 1. This is first seen through the prominent divergence between kinetic and cyber operations specific to the domain of threat. Kinetic conflict has occurred for centuries within the immutable physical world where threat characteristics are physically observable through direct (visual observation) or augmented (technology assisted) means. However, unlike this kinetic conflict situation, the cyber domain is highly malleable and prone to deception. For instance, a spoofing attack is a situation in which one person or computer program successfully masquerades as another by falsifying data and thereby gaining an illegitimate advantage (Gantz & Rochester, 2005).

Further, classic military doctrine in which the defender has numerous advantages (e.g. defensive fortifications and advantageous information asymmetries) is completely up-ended in the cyber domain where the attacker is advantaged. The advantages to the cyber-attacker are numerous and include: (1) anonymity: the ability to hide in a global network across national sovereignty and jurisdiction boundaries complicates attack attribution, (2) targeted attacks: adversaries can pick the time, place, and tools, (3) exploitation: global reach to probe weaknesses of the cyber-defense, (4) human weaknesses: trust relationships are susceptible as evidenced in "social engineering" attacks, and (5) forensics: volatile and transient nature of evidence complicates attack analysis, which can be quite cumbersome (Jain, 2005). Although there are differences between kinetic and cyber domains, it is likely that many of these challenges to cyber operations can be addressed by applying lessons learned from the successful management of kinetic operations.

*Table 1. Comparison of Domain Characteristics of Kinetic and Cyber Operations*

|  | **Kinetic Operations** | **Cyber Operations** |
|---|---|---|
| **Domain** | Physical world that is largely immutable | Virtual world that is highly malleable and prone to deception |
| **Military Doctrine** | Defender has the advantage | Attacker has the advantage |
| **Mathematical Definition** | Force-on-force engagements defined by Lanchester equations[1] | Cyber engagements are potentially scale-free [2] |
| **Requisite Resources** | Resource intensive, requiring an organization with integrated capabilities (i.e. logistics) | Resource non-intensive; scales down to an individual with few prerequisite capabilities |
| **Threat Characteristics** | Physically observable | Hidden in network, can take on many forms |

| **Massing of Forces** | Space and time dependent, advance warning possible | Unconstrained, with little to no advanced warning |
|---|---|---|
| **Detection** | Distributed with many possibilities "hits" from sensors to ISR[3] assets to patrolling units. | Dependent on automation and rule-based Intrusion Detection Systems (IDS); analyst combs through huge log files and creates new attack signatures |
| **Big Data Challenge** | Managing data collection. Intelligence analysis. | Detection (attack correlation) Forensic analysis |
| **Analytical Challenge** | Finding insurgent networks among populace | Finding new threats / developing signatures |
| **Attack Characteristics** | Unfold in space and time | Instantaneous and massively parallel |
| **Effects** | Linear effects whose impact is known, immediate, and attributable to an adversary | Non-linear (potentially cascading) effects whose impact may be hidden, unknown, undetected for periods of time, and non-attributable to an adversary. |
| **Battle Damage Assessment** | Observable and quantifiable | Some observables with many potentially complex higher-order effects; requires forensics and can take months. |
| **Visualization** | Common operational picture; counter-insurgency operations require network analyses and dependency graphs | Attack graphs, dependency graphs, and cyber terrain[4]. |
| **Deception** | Largely at the strategic level, requires substantial planning and is resource intensive | Largely at tactical levels, requires little planning and is not resource intensive |

[1] Bowen & McNaught (1996), [2] Moffat (2006) [3] Intelligence, Surveillance and Reconnaissance, [4] Jakobson (2011)

## 1.2 The Importance of Situation Awareness

It is likely that the dynamics of conflict are extensible to the virtual battlefield. Some key concepts with which to compare kinetic and cyber conflicts are derived from a conceptual framework of *network-enabled operations* underlying information-age conflict (Alberts, 2002; Alberts, Garstka & Stein, 1999). This framework is comprised of four primary tenets (Alberts & Hayes, 2003):

1) A robustly networked force improves information sharing and collaboration
2) Such sharing and collaboration enhance the quality of information and shared situational awareness
3) This enhancement, in turn, enables further self-synchronization and improves the sustainability and speed of command
4) The combination dramatically increases mission effectiveness

Many of these payoffs to network-enabled operations are conceptualized at human and organizational levels in terms of maintaining and enhancing SA, which can in turn lead to force-synchronization and increased mission effectiveness. This conceptual framework explicitly assumes that greater information sharing in a networked organization produces better SA. SA is defined as "the ability to maintain a constant, clear mental picture of relevant information and the tactical situation including friendly and threat situations as well as terrain" (Dostal, 2007). We subscribe to a theoretical model of SA described by Endsley (1988; 1995) in which SA is the perception of relevant elements (e.g., status, attributes, dynamics) in the environment within a volume of time and space (Level 1), the comprehension or understanding of their meaning (Level 2), and the projection of future actions (Level 3).

The tenets of network-enabled operations are posited to yield cumulative effects to organizational effectiveness in military conflicts. Performance and effectiveness may be limited by a failure or bottleneck at any step in the sequence. For instance, an increase in information available to commanders and their staff is postulated to increase the quality of decision-making due to enhanced SA. There may be situations, however, where increased information sharing increases the *quantity* of available information without a corresponding increase in *quality*. The sheer volume and rapid pace of information received and readily accessible through networked systems can be overwhelming. This presents a challenge to the command staff as there are clear limits to human cognition and how much information can be attended to, processed, and shared in a given amount of time, which can potentially limit situational awareness. The following subsections highlight the importance of SA to conflict-based situation management across kinetic (conventional) and cyber (virtual) battlefields.

**1.3 Kinetic SA**

On the conventional battlefield, information is largely gathered directly whether by physical sensors, human sensory perception, or tele-operation of unmanned intelligence, surveillance, and reconnaissance platforms. This corresponds to Level 1 situation awareness (SA). The battlefield is physical and immutable and the opposing forces perceive various states of the same physical battlefield and have access to many similar elements of situational information. In kinetic operations, SA is often dependent upon a careful analysis of the geography of the physical terrain (major waterways, roads, etc.) coupled to target sightings and movements, and friendly positions. Developing and maintaining an accurate *analog* model of the physical battlefield is a critical process, whether "sand tables" (prior 1960s), board game varieties (1960-1980s), or maps with digital overlay (since 1990s). Such models are critical for both perception of the battlefield and comprehension by reasoning about it.

However, there is often a tradeoff between data acquisition and comprehension. Additional efforts in data acquisition may provide more information about the battlefield space; however, adding too much data could overwhelm human processing capacities to analyze the information in a timely manner, greatly impacting comprehension of the current situation. A key research question is understanding the limits to human information processing and how they are manifest in complex, information-rich and broadly-collaborative networked operational environments.

**1.4 Cyber SA**

Technological advancement of the information-age continues to push us towards virtual conflict of networked organizations and individuals. Through mediums such as the Internet, traditional geographical boundaries are subsumed. Thus, a primary goal on the virtual battlefield is to mount a robust cyber-defense. Cyber analysts clamor for advanced capabilities to support their cyber mission and provide better SA. These should include the capabilities that automatically map all paths of vulnerability through networks; correlate and fuse data from a variety of sources; provide visualization of attack paths; automatically generate mitigation recommendations; and ultimately produce analysis of mission impact from cyber attacks (Jajodia et al, 2011).

## 2. Examples of Research in KSA

In the following sections, we describe a challenge in kinetic warfare, and then attempt to review what is known about related challenges in cyber world. In some cases, we find significant similarities or at least parallels, while in others we find instructive differences. In yet other cases, too little is known yet about challenges – or lack thereof – in cyber situation awareness (CSA), and therefore in such cases we merely point out a potential research direction. We begin by describing two examples of research efforts that quantified and illustrated significant aspects of KSA, particularly the characteristic challenges of KSA experienced by practitioners.

**2. 1 Example of Research in KSA:  The DARPA MDC2 program**

The first of the two examples of KSA research we use in this chapter is the program called Multicell and Dismounted Command and Control (Kott, 2008), performed in 2004-2007 by the United States' Defense Advanced Research Programs Agency (DARPA). The main thrust of that research was an experimental exploration of battle command in light-armored, information-rich, distributed networked forces. At the time, the U.S. Army was eyeing the possibility of future combat force based on a combination of fighting units mounted on fast-moving, lightly armored vehicles with large number of sensors – flying drones and autonomous ground sensors – and precise, far-shooting weapons. Such combat units would rely far less on the thickness of their armor than today's ground forces, and far more of their ability to see and destroy the enemy from far away.

In effect, in such a concept, the combat unit was trading the value of heavy armor for the value of advanced information about its enemy. The concern with this concept was whether the human soldiers, the consumers and users of all the rich information that would enable the defectiveness of this hypothetical combat force, would be able to absorb, comprehend and act on this complex and voluminous information. In other words, whether the cognitive challenges imposed by information-rich command environment would prove to be insurmountable.

Since previous battle command systems were not designed to function in such an information-rich environment, a prototype of a new human-machine system was created to translate high-rate inflow of battlespace data into high-quality situation awareness and command decisions. The new prototype system included specially developed situation awareness tools that continuously and autonomously fused all data into a shared situation portrait. Also included were action execution tools that helped human soldiers control intelligence gathering, movements on the battlefield and assessments of results of long-range attacks by precision weapons. Here we start to see some similarities to CSA – very large volume of information and relative absence of direct physical cues.

The research proceeded through a series of intricately organized and rather expensive experiments--simulated battles. In each battle, the Blue Force were U.S. Army soldiers who sat in mock-up battle vehicles equipped with elaborate information systems, and fought a reasonably realistic battle against the well-trained Red Force, portrayed by military professionals. The battle was fought on a simulated battlefield where special simulation software calculated and depicted all physical effects – movements of vehicles, observations of sensors, and shooting of weapons. A set of instrumentation and human observers recorded the state of situation awareness, including the degree of awareness that could be potentially possible given the available information, and the degree of awareness actually exhibited by the soldiers. The actual state of the battle was also recorded for every moment of the battle, e.g., how many Red soldiers were in a particular forest, as opposed to what Blue sensors observed or Blue soldiers recognized. This allowed quantitative tracking of situation awareness overtime, using metrics that combined location, state of health, priority and quantity of opponent's forces. Such metrics could be analyzed also by comparing them with soldiers' understanding of the available information as transpired form their verbal exchanges and actions. In a later section, we continue to discuss the KSA-related findings of this program in comparison with Cyber SA.

## 2.23. Another Example of Research in KSA: the RAID Program

The research program titled Real-time Adversarial Intelligence and Decision-making (RAID) was sponsored by the United States' Defense Advanced Research Programs Agency (DARPA) during the period of 2004-2008 (Kott, 2007; Kott et al., 2011;Ownby and Kott, 2006). The objective of the program was to build tools for automated generation of enemy situation estimates and predictions of enemy near-term action (Level 3 SA) in military operations. A part of the program was also to measure the situation awareness of the human soldiers and to compare their awareness with the estimates of the automated tool.

The RAID program focused on an intentionally narrow but still very challenging domain: tactical combat of Blue Force (infantry, supported by armor and air platforms) against the Red Force (an insurgent-like irregular infantry) in an urban terrain. The problem situation may involve the defense of Blue facilities, the rescue of downed aircrew, the capture of an insurgent leader, the rescue of hostages or the reaction to an attack on a Blue patrol.

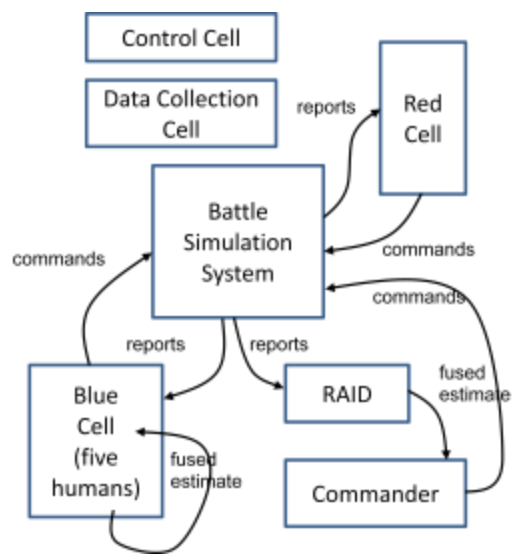

Figure 2. Formation of KSA in the RAID system.

In planning and executing a battle like this, the company commander, his supporting staff (including possibly the staff at the higher echelon of command) and his subordinate unit leaders would receive and integrate (mentally or with the aid of computerized fusion system like RAID) a bewildering array of information (Fig. 2). For example, information on the Blue force composition and mission plan; detailed maps of the area, potentially including detailed 3D data of the urban area; known concentrations of non-combatants such as markets; culturally sensitive areas such as worship houses; reports of historic and recent prior activities such as explosions of roadside bombs in the area; continuous updates on the locations and status of the Blue force as they move before and during the battle.

Using all this information, commander and staff typically produce two types of output. First is the estimate of the Red force's current situation: estimated actual locations of the Red force (most of which are normally concealed and are not observed by the Blue force); the current intent of the Red force, and potential deceptions that the Red force may be performing. The second type of output describes the estimated future events: Red force's future locations (as a function of time), movements, fire engagements with the Blue force, changes in strengths and intent.

Each of multiple experiments in the RAID program consisted of wargames executed by live Red and Blue commanders in a simulated computer wargaming environment. In half of the wargames the Blue commander received the support of a human team of competent assistants (staff). Their responsibilities included producing estimates of enemy situation. This set of wargames constituted the control group. In the other half of wargames Blue commander operated without a human staff. Instead, he obtained a similar support from the RAID automated system which produced enemy situation estimates. These wargames constituted the test group. The data collection and redaction process compared the accuracy of the control group with the accuracy of the test group. In effect, we were able to compare situation awareness of human staff with that of the automated tool. Like the MDC2 program, the RAID program also yielded observations about KSA which we will compare with those of CSA in the following section.

## 3. Instructive Similarities and Formidable Differences between KSA and CSA

We now turn to selected experimental findings of the two KSA-focused programs introduced above, and compare them to those in CSA. Challenges and opportunities of KSA and CSA are similar or at least parallel in several important ways. In both kinetic and cyber worlds, SA strongly impacts the outcome of the mission. The process of developing effective and efficient SA through information collection (Level 1 SA), organization (Level 2 SA), and sharing (Level 3 SA) is difficult to manage in both KSA and CSA (Kott & Arnold, 2013). Effective SA and concurrent decision making can be limited by an individual's cognitive biases. Collaboration for the sake of forming shared SA is another challenge common to both kinetic and cyber worlds. However, the need for collaboration is often a requirement of kinetic conflict, while cyber conflict is often managed at the individual level. Further, collaboration itself is often difficult, particularly because cyber defenders of different roles and backgrounds do not yet share a common set of concepts, terms and boundary objects. These are yet to emerge in this young field. Also, more than in the kinetic world, cyber defenders may need stronger tailored pictures of the same shared model. Table 2 highlights the key similarities and differences that are further discussed in the sections below.

*Table 2.*

| | KSA | CSA | Research Direction |
|---|---|---|---|

| Mission Outcome | Characterized by quantitative, tangible metrics (e.g., location and number of enemy targets) | Mission-defined metrics are not well understood | Development of CSA metrics related to mission and mission outcome |
| --- | --- | --- | --- |
| Representation | Tends to have a commonly accepted, widely used organizing paradigm – the physical terrain of the battlefield (e.g., map) | No map-like common reference has emerged | Development of a shared non-physical network "map" |
| Information Collection, Organization, and Sharing | Challenged by difficulties with timely processing of large amounts of data about current battle state space (e.g., managing dynamic, moving, and relatively scarce sensors) | Challenged with the organization, coordination, and timely analysis of volumes of heterogeneous information from automated sensors, intrusion detection systems, and correlating analytical reports | Approaches to effective representation and fusion of information at optimal levels of abstraction |
| Cognitive Bias | Largely suffer from confirmation bias and availability heuristic | Some evidence of risk-aversion as well as confirmation bias and availability heuristic | Additional research on the formation and mechanisms related to cognitive bias |
| Collaboration / Shared SA | Collaboration has to be controlled, encouraged, and synchronized in order to mitigate potential staff tendency to aggravate cognitive biases and to misdirect precious cognitive resources | Task responsibilities are managed at the level of the individual and are often not shared | Given the malleable nature of the cyber domain, a common set of concepts, terms, and boundary objects are critical for developing CSA and should be a priority for research |

## 3.1. KSA and CSA Strongly Impact Mission Outcome

The first finding may seem obvious – higher KSA leads to notable increase in mission outcome, such as fewer casualties in the battle as compared to opponent's, or the ability to capture the opponent territory or to defend one's own ground. In fact, it is not an obvious finding, and certainly not a well-quantified one in prior work. It becomes particularly non-trivial when we note the difference between the information available to the soldiers and the level of its comprehension, i.e., the cognitive component of situation awareness. On a more fundamental level, one might wonder whether the intangible benefits of higher SA can possibly compare with mighty effects of such tangible factors as speed and armor of combat vehicles, or range and precision of weapons.

Nevertheless, quantitative experimental findings of MDC2 program were unequivocal – higher SA does translate into significantly better battle outcome. Indeed, it was the difference between the amount of information available to the Blue Force versus the information available to the Red Force (the information that is obtained from various sources such as sensors or scouts, and made available to the commander and staffs) that emerged as a key predictor of battle outcome. Because this difference was so important, the Blue force found (empirically) that limiting Red's ability to see the Blue force was critical to winning the battle. The information available to Red routinely increased during the fight when distances between Red and Blue were small enough that relatively weaker Red sensors became effective. As the Blue detection of the Red's high-priority targets increased, so did the potential for battle outcomes favorable to Blue.

Similarly, In RAID program, we also found clear statistical evidence that with more accurate estimates of the Red situation (Level-2 SA, comprehension) and intent (Level-3 SA, projection), the Blue commander was more likely to achieve better battle outcome. To measure the accuracy of SA objectively, we used a metric similar to Circular Error Probable (CEP). Roughly speaking, it gives a typical error between the actual location of an opponent entity, and the location as perceived or projected by the Blue commander. The experimental data were very clear – battle outcome (wargame score) improves as the situation assessment becomes more accurate (i.e., CEP decreases).

In literature on cyber security research, there is recognition, but not yet a quantitative evidence of the impact of CSA on metrics of effectiveness and mission outcomes, such as timely detection if a cyber intrusion. Situation awareness is recognized to be limited in the cyber domain: inaccurate and incomplete vulnerability analysis is common as is the failure to adapt to evolving networks and attacks; and the ability of cyber defenders to transform raw data into cyber intelligence remains quite restricted. Cyber researchers argue that advanced capabilities are needed for mission-centric CSA. These should include the capabilities that automatically map all paths of vulnerability through networks; correlate and fuse data from a variety of sources; provide visualization of attack paths; automatically generate mitigation recommendations; and ultimately produce analysis of mission impact from cyber attacks (Jajodia et al., 2011). Such tools would increase CSA, arguably yielding better cyber defense outcomes, as is the case with KSA.

**3.2 Cognitive Biases Limit Comprehension of Available Information**

Both KSA and CSA may suffer from cognitive biases. The exact manner in which a cognitive bias influences the formation of SA remains a topic for research, in both cyber and kinetic worlds. It cannot be excluded that CSA suffers from different biases than KSA, and perhaps through different mechanisms. In kinetic battles (as found in the MDC2 program), commander and staff surprisingly often dismissed or misinterpreted the available correct information. They also overestimated the completeness and correctness of their KSA, perhaps partly because the advanced sensors and information displays lulled them into false sense of security -- "I can see it all." There was an alarming gap between information available to the commander and staff and the KSA they derived from that information: commander's assessment of the available information was correct only approximately 60% of the time. A cognitive bias – a kind of "belief persistence" – appeared to be a common cause of this inadequacy of comprehension of available information.

In particular, such seeing-understanding gaps often manifested themselves in poor synchronization of information and movements. Commanders frequently over-estimated the strength of the threat they faced, or significantly underestimated that strength. Over-estimate of threat resulted in unnecessarily slowing down the advance of the force in order to acquire more information. Under-estimate of threat caused the force enter into the close contact with the enemy while lacking sufficient information and thereby making Blue Force more vulnerable.

In the RAID program, we also found that human KSA was significantly lower or less accurate than what was possible to achieve using all the available information. We compared two assessments of enemy situation and intent: one produced by humans and another one produced by an automated tool. The tool lacked either the experience or intuition of human soldiers. Nevertheless, the error of the tool's estimates was significantly (on average) lower than that of humans. The fact that the automated tool compared well with competent human staff implied that sufficient information was indeed present in the data, but not extracted by staff in order to yield the best possible KSA.

But why were humans' estimates less accurate than tool's? On one hand, SMEs and psychologists found many similarities in reasoning of humans and of the tool. The difference, however, seemed to be mainly in human cognitive biases. Although such biases are often useful as convenient shortcuts, on balance they lead to reduced accuracy. For example, we often observed a fixation on a presumed pattern or rule: humans estimated the Red situation by applying a previously learned pattern or doctrine. Then, if disconfirming evidence arose, the humans discounted the evidence. When faced with a clever, rapidly innovating Red opponent, such a fixation on previously observed patterns often produced gross errors in Blue's KSA.

Cognitive biases are also widely evident in CSA. Researchers note that cyber defenders exhibit over-reliance on intuition: with few reliable statistics on cyber attacks, decision makers rely on their experience and intuition, fraught with cognitive biases. Such biases are likely to lead to suboptimal decisions. For example, when faced with the trade-off between a certain loss in the present (e.g., investing in improved security) and a potential loss in the future (consequences of a cyber incident), a common risk-aversion bias is toward the second option. In a related observation, cyber defenders tend to believe that their particular organization is less exposed to risks than other parties, particularly if they feel like having a degree of control over the situation-- some refer to this as "optimistic bias." Many are more afraid of risks if they are vividly described, easy to imagine, memorable, and they have occurred in the recent past (related to what is called "availability bias").

Further, a common bias is to ignore evidence that contradicts one's preconceived notions, i.e., the confirmation bias (Julisch, 2013). With respect to the optimistic bias mentioned above, it is important to note that individuals distinguish between two separate dimensions of risk judgment-- personal level and societal level. Individuals display a strong optimistic bias about online privacy risks, judging themselves to be significantly less vulnerable than others to these risks. Internal belief (perceived controllability) and individual difference (prior experience) significantly modulate optimistic bias. (Cho et al., 2010). There is a tendency for individuals to interpret ambiguous information or uncertain situations in a self-serving direction. Perceived controllability and distance of comparison target influence this tendency. (Rhee et al., 2012)

In summary, although cognitive biases play important roles in both KSA and CSA, the limited available literature does not allow us to determine the degree of similarity in specific mechanisms involved. Research in CSA may benefit from an explicit and systematic investigation of whether the biases affecting KSA also play a key role in CSA.

**3.3 Information Collection, Organization and Sharing is Difficult to Manage**

Effective situation awareness takes us through a three-phase process of perception of the data collected (Level 1 SA), organizing said data in a way that it becomes useful information (Level 2 SA), which in turn allows us to make and share decisions based on future predictions (Level 3 SA). Yet this process of information collection, organization, and sharing is difficult to manage in both kinetic and cyber conflicts.

For example, in MDC2 experiments, the commander and staff had difficulties tracking the extent and timing of the sensor coverage available in different parts of the kinetic battlefield. In effect, they often did not know what they had seen and what they had not – an inadequate SA of their own information collection assets. Flaws in sensor layering also led to critical gaps in sensor coverage, which commonly went unnoticed by commanders. These gaps were directly tied to poor KSA related to threat location and proximity, and thus increased the likelihood for encountering an ambush by the opponent.

Especially difficult was management of multiple diverse sensors with significantly different capabilities. Not only they differed in capabilities, area of coverage, agility and latency of information, but also in their organizational ownership and rules of who and when was allowed to use or reposition them. As a result, soldiers had to dedicate a large fraction of their available time and attention to issues of information acquisition. In many cases it became the primary concern of the commander who focused on managing sensor assets, and delegated other tasks. Indeed, over 50% of all decisions were made to acquire information. "Seeing" was considered the hardest task while "shooting" was considered the easiest task. Commander and Staff also found that battle damage assessment has grown as a critical and most demanding task and a key detriment to KSA. Difficulties in assessing the "state" of engaged targets significantly degraded the level of KSA.

Indeed, a major tenet of the U.S. Office of Secretary of Defense's "data to decisions" initiative and a primary challenge for military commanders and their staff is to shorten the cycle time from data gathering to decisions (Swan & Hennig, 2012). A key information-age challenge is that the sheer volume of information available constrains military decision-making cycles, so that the staff is stuck in observe-orient, the "seeing" part of the cycle, rather than advancing further into the decide-act, or "shooting" part of the cycle.

These challenges in KSA parallel the challenges of managing information for CSA. Lack of information, such as reliable statistics on the probability and impact of cyber attacks, induce decision makers to rely unduly on their experience and intuition. In acquiring information about the cyber environment, important classes of information include: (a) the probability of particular types of cyber attacks; (b) the effectiveness of existing countermeasures in defending against these attacks; and (c) the impact or cost of attacks (Julisch, 2013). Because dynamic cyber intelligence is difficult to acquire, over-reliance on static knowledge versus dynamic intelligence is common.

Other peculiarities of cyber security world add to the complexity of information acquisition, management, and related formation of CSA. Missions are generally defined in terms of abstract resources and not actual systems and devices (making comprehension of relations between missions and tangible systems more difficult); organizations often outsource parts or all responsibility for cyber defense (thereby complicating understanding of responsibilities and correlation of information); resources are managed in a highly dynamic fashion; and increasingly large number of sensors overload human analysts (Greitzer et al., 2011).

Cyber researchers note additional related challenges: information sharing methods are immature, especially as the process of forming CSA is distributed across human operators and technological artifacts operating in different functional areas. Add to this the rapid rate of environmental change, overwhelming volume of information, and lack of physical world constraints. With such ensemble of challenges in information acquisition and management, it is not surprising that CSA is distributed, incomplete, and domain-specific (Tyworth et al; 2012).

## 3.4. Collaboration can be Challenging

The essence of collaborative teamwork within an organization is the ability to efficiently maintain a coherent set of tasks across multiple people and shared assets. It is commonly understood that SA benefits from effective collaboration of participants of the SA-generation process. However, collaboration also can have a dark side and exact a high cost. In MDC2 experiments, we observed on a number of occasions that a commander's KSA degraded as a result of collaborations with subordinates, peers, or higher echelon decision-makers. Collaboration can reinforce an incorrect perception by apparent acquiescence by other decision makers. Information gaps – the importance of which we mentioned earlier in this chapter – are not necessarily appreciated by individual commanders, and collaboration does not help to alleviate that.

As an example, out of seven episodes of collaboration in a particular experiment in the MDC2 program, three episodes produced improved KSA, two collaboration episodes distracted the decision-maker from the more critical focus, and two others led the decision-makers to reinforce the wrong interpretation of the situation. The mechanisms by which collaboration may impose costs on KSA vary: in some cases collaboration tends to reinforce confirmation bias; in other cases collaboration mis-directs the attention away from most critical issues.

In the RAID program, we observed a negative correlation between the number of collaboration events within the staff, and the quality of KSA. This could be explained as follows: more intensive collaboration may lead to greater consumption of cognitive resources, resulting in lower accuracy of KSA and lower battle score.

In the literature on cyber defense, we do not find concerns about a potential negative impact of collaboration on CSA. However, concerns about the difficulties of enabling effective collaboration are common in the world of cyber defense. On one hand, collaboration in cyber security teams can be very effective. Experiments in synthetic IDS environment demonstrate that collaborative teams outperformed individuals on average. However, this appears to apply when the teams focuses on "hard" cases requiring diverse expertise. It is not unlikely that in "easy" cases, collaboration could be counterproductive (Rajivan et al., 2013).

On the other hand, it is argued that in cyber defense, collaboration suffers from the lack of boundary objects (i.e., intermediate products that can be shared – common in more mature fields of practice). CSA tends to be distributed, incomplete, and highly domain-specific. Boundary objects that have emerged in cyber defense are currently limited to reports; these are inadequate and not as effective as boundary objects in other fields. To alleviate the current lack of commonly understood boundary objects, cyber defense may benefit from visualizations capable of presenting cross-domain information for domain specific purposes (Tyworth et al; 2012).

Other cyber researchers highlight the difficulty of assessing trustworthiness in collaborative communications. For example, means for numeric and verbal communication of cybersecurity risks are not yet adequately developed and are poorly understood in cyber defense (Nurse et al., 2011). This is further exacerbated by barriers between individuals of different roles and backgrounds in cyber defense. For example, cyber experts see users both as potential cyber defense resources, but also as sources of accidents and potential threats. Unlike users, experts tend to use probability rather than consequences as a basis for evaluating risk. In addition, experts' lack of detailed knowledge of their users' information security performance complicates effective collaboration (Albrechtsen & Hovden.,

2009). As a result, CSA suffers.

**3.5 Shared Picture does not Assure Shared SA**

In addition to effective collaboration, shared picture of the situation is often seen as a key to collaborative SA. However, experiments in MDC2 indicated that sharing picture is no substitute for sharing intent. While a commander often thought that his subordinates understood his intent because they could see it all on the screen, the subordinates in fact could not perceive the commanders intent from the picture he shared with them. And when staff members do not share the commander's SA, including an understanding of the commander's intent, they may be less likely to take initiative.

Perhaps this should not be surprising: because different viewers of the same "shared" picture differ significantly in their roles and backgrounds, they should see different, properly tailored pictures in order to arrive to a common SA. Indeed, some cyber researchers argue that the common picture should not be common. Modalities of interaction and information requirements are inherently different for different types of users. One proposed approach is a model-based cyber defense situation awareness: a common model represents the current security situation of all protected resources, updated over time. Based on this common model, different intuitive visualization can be employed for different users (Klein et al., 2010).

# 5. Summary

By exploring what is known about SA in conventional – also commonly referred to as kinetic – battles, we may gain insights and research directions relevant to cyber conflicts. For the sake of brevity, we use the abbreviation CSA for Cyber Situation Awareness and KSA for Kinetic Situation Awareness. The

Information Age is defined by the rise of networked forms of organization and an increase in information available to commanders and their staff is postulated to increase the quality of decision-making due to enhanced situational awareness. However, there are clear limits to human cognition and how much information can be attended to, processed, and shared in a given amount of time, which can potentially limit situational awareness. Challenges and opportunities of KSA and CSA are similar or at least parallel in several important ways. In both kinetic and cyber worlds, SA strongly impacts the outcome of the mission. In literature on cyber security research there is recognition, but not yet a quantitative evidence of the impact of CSA on metrics of effectiveness and mission outcomes. Researchers and practitioners of KSA have a commonly accepted, widely used organizing representation – map of the physical terrain of the battlefield. Yet no map-like common representation has emerged in CSA. It is likely, although not yet examined, that cognitive biases are general to both KSA and CSA. For example, in KSA, the human tendency to look for confirming evidence has routinely been exploited in intelligence deception. Cognitive biases are also widely evident in CSA, such as "optimistic bias." Limited or incorrect incoming data, such as reliable statistics on the probability and impact of attacks (whether kinetic or cyber), induce decision-makers to rely unduly on their experience and intuition. Collaboration also can have a dark side and exact a high cost. Collaboration may reinforce an incorrect perception by apparent acquiescence by other decision makers.